\documentclass[reprint, aps, amsmath, amssymb, twocolumn,  longbibliography]{revtex4-1}
\usepackage{textcomp}
\usepackage{graphicx}
\usepackage[hidelinks]{hyperref}
\usepackage{xcolor}
\hypersetup{
    colorlinks,
    linkcolor={blue!50!black},
    citecolor={blue!80!black},
    urlcolor= blue
}
\begin{document}
\newcommand{\qg[1]}{\textcolor{blue}}

\preprint{APS/123-QED}

\title{Nanofiber based displacement sensor}% Force line breaks with \\

\author{Chengjie Ding$^{1,2}$, Maxime Joos$^1$, Constanze Bach$^{1,3}$, Tom Bienaimé$^1$, Elisabeth Giacobino$^1$, E~Wu$^2$, Alberto Bramati$^1$, Quentin Glorieux$^1$}
\email[Corresponding author: ]{quentin.glorieux@lkb.upmc.fr}
\affiliation{(1) Laboratoire Kastler Brossel, Sorbonne Universit\'e, CNRS, ENS-PSL Research University, Coll\`ege de France}
\affiliation{(2) State Key Laboratory of Precision Spectroscopy, East China Normal University, Shanghai 200062, China}
\affiliation{(3) Vienna Center for Quantum Science and Technology (VCQ), Faculty of Physics, University of Vienna, Boltzmanngasse 5, A-1090 Vienna, Austria}

\date{\today}% It is always \today, today,

\begin{abstract}
We report on the realization of a displacement sensor based on an optical nanofiber.
A single gold nano-sphere is deposited on top of a nanofiber and the system is placed within a standing wave which serves as a position ruler.
Scattered light collected within the guided mode of the fiber gives a direct measurement of the nanofiber displacement. We calibrated our device and found a sensitivity up to 1.2~nm/$\sqrt{\text{Hz}}$.
As an example of application, a mechanical model based on the Mie scattering theory is then used to evaluate the optically induced force on the nanofiber by an external laser and its displacement.
With our sensing system, we demonstrate that an external force of 1~pN applied at the nanofiber waist can be detected. 
%It paves the way to realize optomechanical based research in the future.
\end{abstract}

\maketitle

%\tableofcontents

\section{\label{sec:level1}Introduction}
Position sensing with micro or nano-meter resolution is  widely used both in scientific research and for industrial applications \cite{muschielok2008nano,andrecka2009nano}. 
 In order to probe weak forces the miniaturization of the devices towards the nanoscale is one of the main approaches \cite{de2017universal,de2018eigenmode}. 
Nanowires \cite{feng2007very}, carbon nanotubes \cite{conley2008nonlinear}, and graphene \cite{eichler2011nonlinear} were successfully demonstrated as potential materials for position sensing.
Several hybrid systems were also used including nanomechanical oscillators coupled to a whispering gallery mode resonator \cite{anetsberger2009near}, superconducting microwave cavities  cooled by radiation pressure damping \cite{teufel2008dynamical}, silicon nanowire mechanical oscillators for NMR force sensing \cite{nichol2012nanomechanical} and nanomechanical oscillators combined with single quantum objects \cite{arcizet2011single}.

Pico-meter sensitivity has been achieved using a quantum point contact as a scanned charge-imaging sensor (sensitivity of 3 pm/$\sqrt{\text{Hz}}$ \cite{cleland2002nanomechanical}) or using a nanorod placed within a Fabry-Pérot microcavity (sensitivity of 0.2 pm/$\sqrt{\text{Hz}}$ \cite{favero2009fluctuating}). 

Here, we propose to use an optical nanofiber as a nanometric displacement sensor which is all optical and can operate at room temperature and in free space.
The optical nanofiber is an air-cladding guiding system with subwavelength diameter, which allows light to be transmitted outside the fiber in the form of evanescent field.
In the recent years, the optical properties of nanofibers have been widely used for atomic fluorescence investigation \cite{nayak2007optical}, atom trapping \cite{vetsch2010optical,corzo2016large}, quantum optics and quantum information \cite{yalla2012efficient}, chiral quantum optics \cite{lodahl2017chiral}, and optical resonators  \cite{ding2019fabrication}. 

Interestingly, a nanofiber has also remarkable mechanical properties, including a small mass, on the order of  0.1 $\mu$g/m and a high  sensitivity to vibrations and weak forces.
In this work, we combine the optical and mechanical properties of a nanofiber to create a unique sensing device for displacement measurements and optomechanical applications.
By depositing a nanoparticle on top of a nanofiber and placing this system in an optical standing wave positioned transversely as a ruler, we report an absolute sensitivity up to 1.2~nm/$\sqrt{\text{Hz}}$.
Using the Mie scattering theory, we evaluate the force induced on the nanofiber by an external laser and demonstrate that a sensitivity of 1~pN  can be achieved. 
%Compare to some numbers of other sensors 
%The analysis of mechanical forces with piconewton sensitivity can also be realized by sensing system includes scanning probe microscopy, atomic force microscope(AFM), fluorescence resonance energy transfer(FRET)-based tension sensors, etc.
Force sensing with pico-newton precision can be used as a sensor for optomechanics \cite{larre2015optomechanical}, biophysics investigations \cite{freikamp2016piconewton} and molecule mechanics \cite{finer1994single}.

The article is organized as follows.
We first introduce the experimental setup and describe the calibration procedure used to measure the sensitivity.
We then present the experimental Allan deviation of our device, measured up to 10 minutes with a $\text{100 }\mu \text{s}$ integration time.
In the second part, we present a possible application of our sensor to detect radiation pressure force.
We propose a mechanical model to estimate the displacement of our sensor under an optically induced force, and we show that the full Mie scattering theory must be taken into account for accurate predictions.
Finally we model the two orthogonal polarisations as a signature of the anisotropy of the system.

\section{\label{sec:level2}Measuring the absolute displacement of an optical nanofiber}
%In this paper, we report on the use of an optical nanofiber as a nanometric displacement sensor.
The general idea of our approach is to place an optical nanofiber within a transverse standing wave and to monitor the amount of scattered light in the guided mode of the fiber in order to infer its position.
Optical nanofibers are optical waveguides with a sub-wavelength diameter.
By carefully tapering a commercial single-mode fiber to sub-wavelength diameter, the adiabatic conversion of the HE$_\text{11}$ mode can be achieved \cite{ward2014contributed,solano2017optical,ding2010ultralow}.
In the sub-wavelength region, the optical mode is guided in air and a strong evanescent field is present outside of the nanofiber.
This has the crucial advantage that light can be coupled in and out with a scattering object deposited, in the evanescent field, on the surface of the nanofiber.

\begin{figure}[]
\centering\includegraphics[width=1\linewidth]{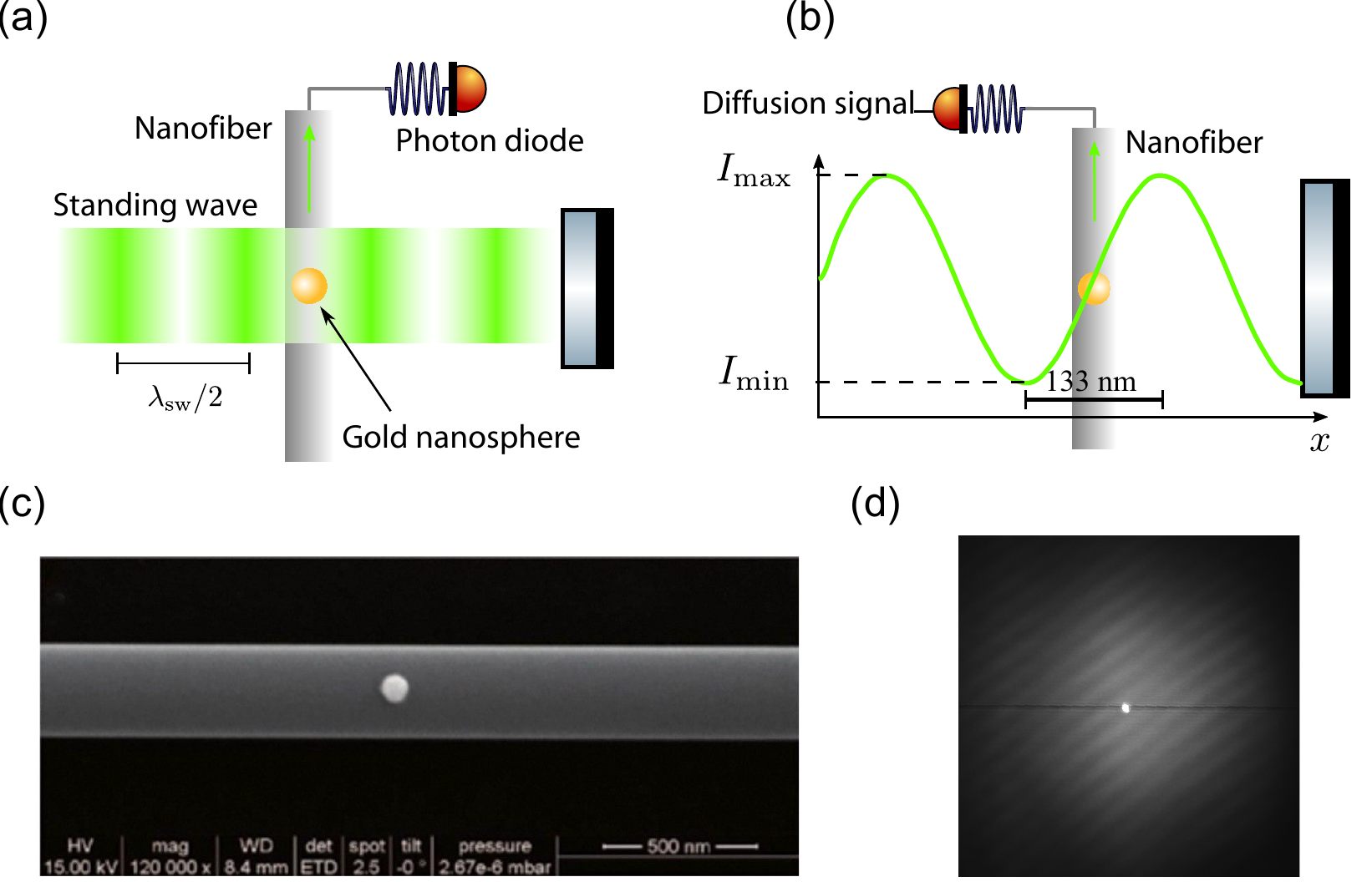}
\caption{a) Sketch of the experimental setup. A gold nano-sphere on the waist of an optical nanofiber is placed in a standing wave ($\lambda_\text{sw}$ is 532 nm). 
The diffusion signal is collected via the fiber.
b) Ruler resolution. Maximum and minimum  intensity are marked as $I_\text{max}$ and $I_\text{min}$, with 133 nm interval.
c) Scanning electron microscope (SEM) image of a single gold nano-sphere deposited on nanofiber.
d) Optical microscope image of a gold-nano-sphere on the nanofiber within the standing wave.}\label{figure1}
\end{figure}

We fabricated a tapered optical nanofiber with a diameter of 300~nm and more than 95\% transmission (at 532 nm), following the method detailed in \cite{ward2014contributed} and used in \cite{joos2018polarization}.
To observe scattering, we overlapped a transverse standing wave with the nanofiber, by retro-reflecting a standard laser diode at 532~nm on a mirror (see Fig.~\ref{figure1}a)).
The mirror is mounted on a piezo-transducer, so that we can scan it in the longitudinal axis to move the standing wave envelope around the nanofiber position. 
In order to guarantee a high contrast of the standing-wave, we verified that the coherence length of the laser is much longer than the distance between the mirror and the nanofiber.
In practice, this is not a stringent condition for standard narrow linewidth laser diode. 

In this configuration the light scattered inside the guided mode of the nanofiber is only due to the rugosity of the nanofiber and is extremely low.
To guarantee a high signal to noise ratio on the detection and therefore a high displacement sensitivity, we have deposited a gold nano-sphere on the nanofiber to enhance scattering (see Fig.~\ref{figure1}c).
The nano-sphere has a radius of 50~nm.
The deposition is done by filling a highly dilute solution of gold nano-spheres inside a micro-pipette and touching repeatedly the nanofiber with the meniscus at the extremity of the pipette.
We continuously inject light inside the nanofiber to monitor the scattered light from the deposition region. 
As seen in Fig.~\ref{figure1}d, the scattered light increases dramatically after the deposition of the gold nano-sphere  \cite{yalla2012efficient}.

The nanofiber is mounted on a custom holder, which has a bending piezo transducer at one side, in order to control its tension.
Finally, we placed the system under vacuum to avoid dust particle deposition and perturbations from air-flows. 
The scattered light guided into the fiber is detected with an avalanche photodiode to have high detection efficiency (50$\%$) and fast response up to 100 MHz.

%The precision is not only related to the “minimum scale of the ruler”, which is decided by the wavelength of the standing wave, but also the size of the "probe", which in our case is the size of the gold nanoparticle.

\begin{figure}[htbp]
\centering\includegraphics[width=1\linewidth]{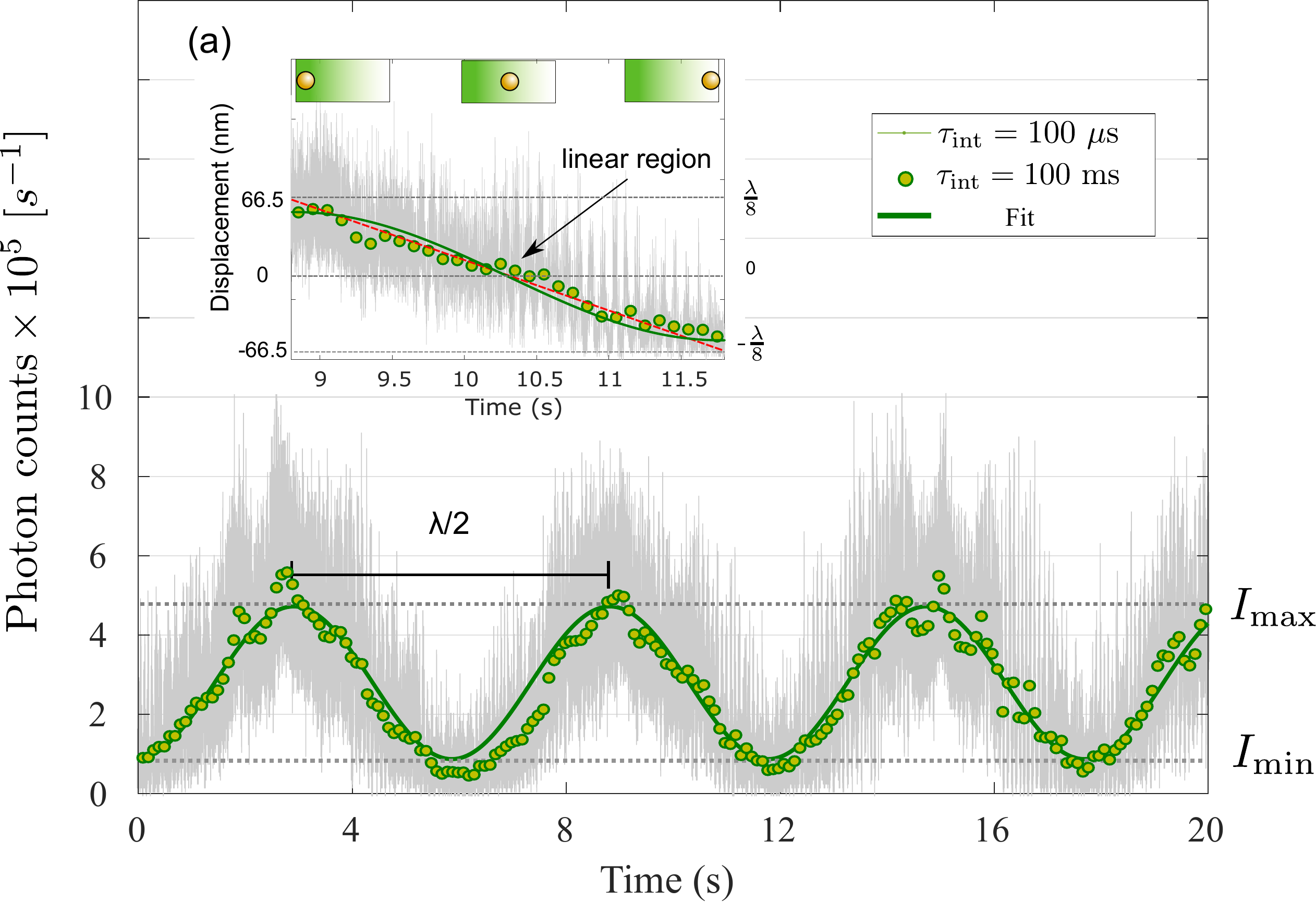}
\caption{Collected photon counts per second while scanning the standing wave.
Integration time is $\tau_\text{int}=100~\mu \text{s}$ for grey lines and $\tau_\text{int}=100$~ms for  green dots.
The solid green line is a sinusoidal fit of the data with $100$~ms integration time to determine the visibility.
$I_\text{max}=4.7$ and $I_\text{min}=0.87$.
Inset: the maximum resolution region is fitted with a linear trend. }\label{figure2}
\end{figure}

To calibrate the displacement, we first scan the standing wave around the nano-sphere.
The signal oscillates with a period corresponding to $\lambda_\text{sw} / 2 = 266$ nm.
The visibility of the standing wave is defined as $v=\frac{I_\text{max}-I_\text{min}}{I_\text{max}+I_\text{min}}$
where $I_\text{min}$ and $I_\text{max}$ are the minimum and maximum detected intensity.

A sinusoidal fit of the experimental data is presented in Fig.~\ref{figure2} and gives a visibility of 0.71 calculated using the the measurement with 100 ms integration time.
In Fig.~\ref{figure2}, we can see the vibrations in the environment through the sensing system.
These vibrations have high amplitude which widened the scan curve.
For this measurement a relatively long integration time (100 ms) has been used and therefore high frequency noise reduces the contrast below 1.
%At shorter integration time, a visibility up to XXX has been measured experimentally.
For small displacements ($d\ll \lambda/4$), the most sensitive region of the standing wave is obviously the linear part (see Fig.~\ref{figure1}b).
In this region, the calibration of the displacement as a function of the scattered light is obtained by fitting the region with a linear model.

%Since our standing wave sensing system is very sensitive, we can see the vibrations in the environment through the sensing system. These vibrations has high amplitude with high frequency which widen the scanned curve. 

\section{\label{sec:level3}Sensitivity of the system}%sensitivity of the system
The sensitivity of our system is linked to two main factors: at short integration time, the noise  essentially comes from the fluctuations of the scattered light and the mechanical oscillations of the nanofiber and, at long time scale, we observed a drift caused by a temperature shift in the environment which displaces the standing wave.

\begin{figure}[htbp]
\centering\includegraphics[width=\linewidth]{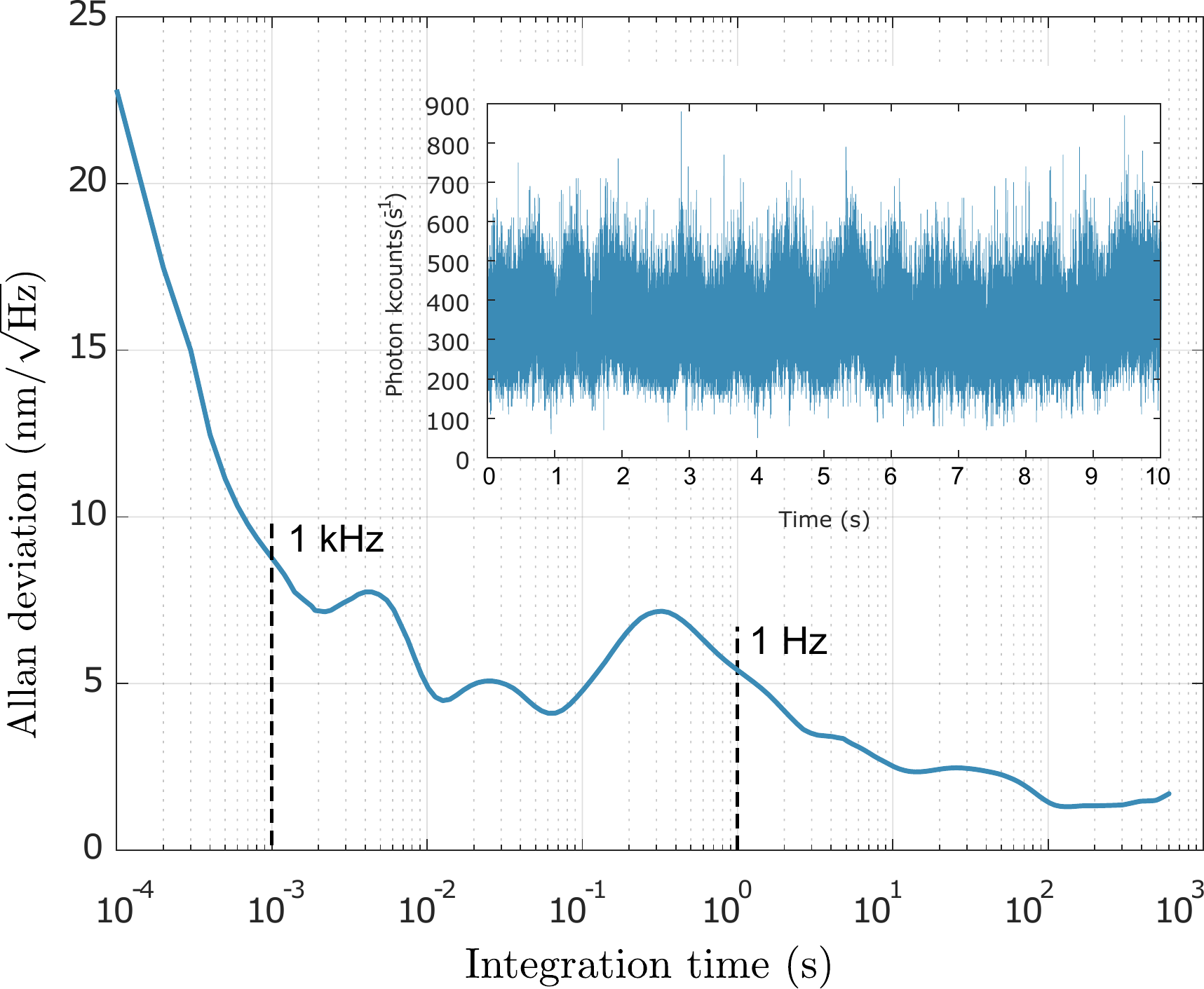}
\caption{Allan deviation.
%$\tau=k\tau_0$ is the averaging time, and $\sigma$ is the calculated Allan deviation at each $\tau$. 
Inset: scattering signal versus time.}\label{figure3}
%\caption{Diffusion signal measured with a single photon detector as a function of time with nanofiber in a vacuum chamber (isolated from the outside airflow). Particle is placed at the linear region of the standing wave.}
\end{figure}

A quantitative description of the sensitivity is given in Fig.~\ref{figure3} by presenting the Allan deviation.
The Allan deviation is typically used to measure frequency stability in oscillators but it can also be implemented in the time domain.
Differently from standard deviation which gives only one value for a given integration time, Allan deviation gives access to the whole noise spectrum at different frequencies detected in our system.
We follow the equation given by \cite{allan1981modified}.
$\Omega(\tau)$ is the measured time history of photon counts with a sample period of $\tau_\text{0}$. 
%The data was then divided into many series with different averaging time.
We divide the data sequence (photon arrival time on the photodiode) into clusters of time $\tau$.
The averaging time is set as $\tau=m\tau_\text{0}$, where the averaging factor $m$ is a group of integers from 1 to $n$. $n=100$ was used in calculation.
The Allan deviation $\sigma(\tau)$ is calculated using averages of the output rate samples over each time cluster. 
\begin{equation}
\sigma(\tau)=\sqrt{\frac{1}{2 \tau^{2}}\langle \left(x_{k+2 m}-2 x_{k+m}+x_{k}\right)^{2}\rangle},
\end{equation}
or in practice:
\begin{equation}
\sigma(\tau)=\sqrt{\frac{1}{2 \tau^{2}(N-2 m)} \sum_{k=1}^{N-2 m}\left(x_{k+2 m}-2 x_{k+m}+x_{k}\right)^{2}}
\end{equation}
where $x(t)=\int^{t} \Omega\left(t^{\prime}\right) d t^{\prime}$ is the integrated photon counts. $N$ is the sample points. 

According to the fit in Fig.~\ref{figure2}, we got the correspondence between fiber displacement and photon count, which is about 2.9 kcounts/nm.
Therefore, the Allan deviation is given in unit nm/$\sqrt{\text{Hz}}$ in Fig.~\ref{figure3}.

On Fig.~\ref{figure3}, three main regions can be identified: i) at  integration time shorter than 1~ms (frequency higher than 1~kHz) where the sensitivity increases with the integration time (lower $\sigma$), ii) at integration time between 1~ms and 1~s (between  1~kHz and 1 Hz) where we observe a sinusoidal noise with characteristics bumps in the Allan deviation signature of the mechanical modes of the nanofiber, and iii) at integration time longer than 1~s (frequency lower than 1~Hz) where the sensitivity increases again.
Finally, at longer integration times than 3 minutes (not shown), the sensitivity decreases again (higher $\sigma$) due to the thermal drift in the standing wave phase.
We can extract two key figures of merit from these data. 
The resolution at 1~s integration time is 5.1~nm/$\sqrt{\text{Hz}}$, and the maximum absolute resolution is 1.2~nm/$\sqrt{\text{Hz}}$

\section{\label{sec:level4}Displacement of the nanofiber driven by externally applied force}
In the second part of this paper, we propose one possible application of our nanofiber displacement sensor for optomechanics experiments.
We assess the possibility to detect the radiation pressure force, by estimating theoretically the displacement of the nanofiber when it is driven by an externally applied force and we investigate in particular optically-induced forces based on the Mie theory.
%\textcolor{red}{It's for studying the possibility of detecting radiation force with our nanofiber based displacement sensor.}

We model the nanofiber as an elastic string of circular section (radius $a$) with the two ends fixed.
The location of the applied force on the nanofiber is an important parameter for the sensitivity to the force.
Here, we assume that the force is applied at the middle of the nanofiber waist, i.e. in the most sensitive region, however the model can be extended easily to other positions.
First, we take the initial tension on the nanofiber to be zero.
The expression of the displacement as function of the applied force $F$ is then  given by:
\begin{equation}
\delta=\left[\frac{F}{8 \pi a^{2} E_Y}\right]^{1 / 3} L,
\label{eq1}
\end{equation}
where $E_Y$ is the Young modulus of silica, and $L$ is the length of the nanofiber region \cite{bellan2005measurement, timoshenko1968elements}.

It is interesting to evaluate if a radiation pressure force applied by an external \textit{pushing} laser could be observed using this system with the sensitivity measured earlier.
A coarse evaluation can be made using a simple model by considering the nanofiber with radius $a$ as a beam with squared section $(2a)^2$ and estimate the force by the differential Fresnel reflections on the sides of the silica beam.
In this geometrical approach, the pushing laser is described by its diameter $d$, its power $P$ and its intensity $I = P/(\pi d^2/4)$.
If $d \gg 2a$, the illuminated section of the nanofiber is $2a \times d$ and the force is 
\begin{equation}
    F_{\text{og}} = \frac{2 I}{c}\ 2R\ 2ad = 8\frac{ P}{c \pi d}\ 2R\ 2a,
\end{equation}
with $c$ the speed of light in vacuum and $R$ the normal incidence reflection coefficient.
With this model, we can estimate the force of a flat top beam of $P = 100\text{ mW}$ and diameter $d=10\text{ }\mu\text{m}$ on a \textit{nanofiber} of squared section $2a = 500\text{ nm}$ made in silica (optical index of $1,5$ which induces a Fresnel reflection coefficient $R = 0,04$):
\[
F_{go} \sim 3 \text{ pN}.
\]

However, if this simple model is instructive to obtain an order of magnitude of the applied force, it must be refined to take into account the interference and polarization effects.
In the following, we use the Mie theory of diffusion to give a more precise estimation of the optical force.
We assume a cylindrical nanofiber (radius $a$ and index of refraction $n_1$) illuminated on normal incidence by a plane wave (with Gaussian envelope of width $d$). 
%This configuration is represented in Fig.~\ref{figure5}.

\begin{figure}[htbp]
\centering\includegraphics[width=0.85\linewidth]{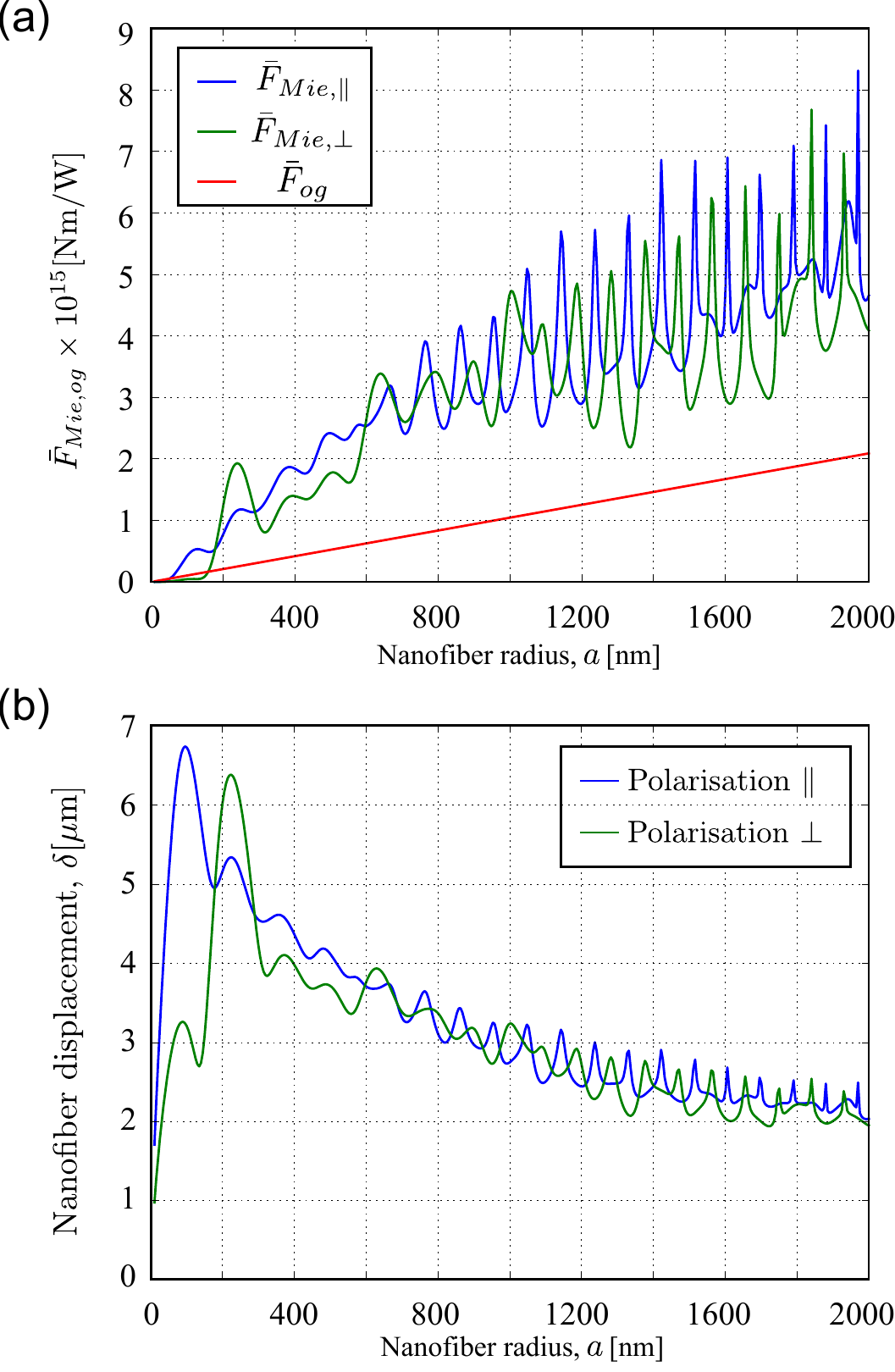}
\caption{Estimation of the radiation force on a silica nanofiber.
(a) Radiation force $\bar{F}_\text{rad}$ on a nanofiber per unit of intensity and per unit of length of illuminated fiber as a function of the nanofiber radius.
(b) Deflection of the nanofiber under the effect of the radiation force $\delta$ as a function of nanofiber radius. 
Parameters: flat-top laser beam ($\lambda$ = 780 nm), with power $P$ = 100 mW and diameter $d$ = 10 $\mu$m, the Young's modulus of silica $E_\text{Young}$ = 70 GPa and the length of nanofiber $L_\text{0}$ = 10 mm.}\label{figure5}
\end{figure}
Two linear polarisations can be used, either parallel or perpendicular to the nanofiber axis.
The force given by the Mie theory is  \cite{mitri2017radiation,loo2019imaging}:
\begin{equation}
    F_{\text{Mie}} = \frac{2I}{c} a d Y,
    \label{eq:Frad_Mie}
\end{equation}
where $Y$ is a dimensionless quantity which depends on the polarisation.
For a linear  polarisation parallel to the nanofiber $Y = Y_{||}(n_1, k, a)$ is given by
\begin{equation}
    Y_{||}(n_1, k, a) = -\frac{1}{ka}\text{Re}\left[ \sum_{l=-\infty}^{+\infty} b_l(b_{l+1}^* + b_{l-1}^* + 2)\right].
\end{equation}
The geometric coefficient $b_l = b_l(n_1, k, a)$ is given by 
\begin{equation}
b_l = \frac{n_1J'_l(n_1ka) J_l(ka) - J_l(n_1ka) J'_l(ka)}{n_1J'_l(n_1ka) H^{(1)}_l(ka) - J_l(n_1ka) H'^{(1)}_l(ka)},
\label{eq:Mie-coeff}
\end{equation}
with $H_l^{(1)}$ is the first order Hankel function and  $J_l$ is the first order Bessel function.
Similarly, for a linear polarisation perpendicular to the nanofiber, we have
\begin{equation}
    Y_{\bot}(n_1, k, a) = -\frac{1}{ka}\text{Re}\left[ \sum_{l=-\infty}^{+\infty} a_l(a_{l+1}^* + a_{l-1}^* + 2)\right],
\end{equation}
with
\begin{equation}
a_l = \frac{J'_l(n_1ka) J_l(ka) - n_1J_l(n_1ka) J'_l(ka)}{J'_l(n_1ka) H^{(1)}_l(ka) - n_1J_l(n_1ka) H'^{(1)}_l(ka)}.
\label{eq:Mie-coeff2}
\end{equation}.

This model is compared numerically to the geometric model in Fig.~\ref{figure5}a). 
It can be observed that the geometric model, represented by the red line in the figure clearly under-estimate the force for both polarisations above a nanofiber diameter of 200~nm.
Compared to the geometrical optics model, the Mie scattering model also predicts oscillations in the radiation force as a function of the increasing radius of the nanofiber, which corresponds to an interference between the specularly reflected waves from the edge of the nanofiber and those reflected by its dielectric core.
In addition, the radiation force shows dependence on the incident polarization which is characteristic of the anisotropy of the system \cite{joos2019complete}.

In the following, we use the full model of Mie scattering to estimate the displacement.
We inject the expression of the force in the displacement of the nanofiber given by Eq. (\ref{eq1}) and we trace the displacement as function of the nanofiber radius in the absence of tension $T_0$.
In this configuration, we predict a displacement on the order of 5 $\mu$m for both polarisations at fiber diameter around 400~nm. While $5 \, \mu\text{m}$ far exceeds the linear dynamical range introduced in the first part of this paper, we can easily reduce the power of the beam generating the radiation pressure force such that the beam displacement is compatible with the requisite range of our experimental technique (displacement smaller than 266 nm).
The minimal force that can be detected with our device is then on the order of 1~pN.
Finally, to extend our model, we can include a non-zero initial tension on the fiber.
In presence of an initial tension $T_0$, a correction term can be added to Eq. (\ref{eq1}) to determine the force $F$ required to induce a displacement  $\delta$ of the nanofiber:
\begin{equation}\label{equation6}
F=8 \pi a^{2} Y_E\left[\frac{\delta}{ L}\right]^{3}+4 T_{0}\frac{\delta}{ L}
\end{equation}
The first term is linked to the mechanical properties of silica as the second one is a consequence of the nanofiber tension.
As $\delta / L \ll 1$ in our experiment, tension might play an important role in the ability of our system for detecting weak force applied on the fiber. 
Experimentally, we have anticipated this situation by adding a bending piezoelectric transducer to the nanofiber holding mount.
This open the way to future investigations of the tension as well as the potential role of damping due to air around the nanofiber.

% To realize ultra-sensitive investigations of force, we need to pay attention to the way that we place the nanofiber. Placing the fiber horizontally is a way to reduce the influence of the gravity of the fiber itself acting as tension on the nanofiber. Because when the fiber is placed vertically, the gravity of the bottom part can be compensated by the support force of the holder, while for the higher part of the fiber, the gravity will be compensated as tension. Secondly, instead of a both ends fixed fiber holder, we used a fiber holder which has a bending piezo at one side, so that we can adjust the tension.
% Ideally, we want to remove the tension around the nanofiber region. In this case, the linear term $4 T_{0}(\delta / L)$ in equation (\ref{equation6}) can be neglected.
% \begin{figure}[htbp]
% \centering\includegraphics[width=1\linewidth]{figure4.png}
% \caption{Displacement of nanofiber as a function of force applied.}\label{figure4}
% \end{figure}

% As shown in Fig.\ref{figure4}, we only need force below pN scale to push the fiber about 1 $\mu$ m. In our experiment, we calibrated the precision of our nanofiber sensing system is 8 nm, which means our system is sensitive enough to detect an externally applied force around 1 pN.

\section{\label{sec:conclusion}Conclusion}
In conclusion, we have presented a position sensor based on a gold nano-sphere deposited on a nanofiber and  placed within a standing wave.
Our sensor can measure the displacement of a nanofiber driven by externally applied force and vibrations.
We experimentally calibrated our sensing system and obtained a resolution of 1.2~nm/$\sqrt{\text{Hz}}$.
We proposed a mechanical model to estimate the response of our sensor to optically-induced pressure force and we found that the sensitivity corresponding to the sensibility of an externally applied force at the nanofiber waist is 1~pN.
This optical nanofiber sensor paves the way to realize integrated optomechanical  research using this platform, such as the demonstration of superfluidity of light in a photon fluid \cite{larre2015optomechanical,fontaine2018observation}.

\section*{\label{sec:ack}Acknowledgement}
The authors would like to thank Arno Rauschenbeutel for designing an early prototype of this device and for fruitful discussions on the sensitivity measurements.
This research is supported by the Emergences Ville de Paris Nano2 project, the Caiyuanpei Programm and the European Union’s Horizon 2020 research and innovation program under grant agreement No 828972.
C.D. is supported by a CSC scholarship.
\bibliography{Reference.bib}
\bibliographystyle{unsrt}
\end{document}